# Atomistic Modelling of Functionally Graded *Cu-Ni* Alloy and its Implication on the Mechanical Properties of Nanowires

Md Shajedul Hoque Thakur[1, a)], Mahmudul Islam[1, b)], Nur Jahan Monisha[1, c)], Pritom Bose[1, d)], Md. Adnan Mahathir Munshi[1, e)], and Turash Haque Pial[2, f)]

[1]Department of Mechanical Engineering, Bangladesh University of Engineering and Technology (BUET), Dhaka-1000, Bangladesh
[2]Department of Mechanical Engineering, University of Maryland, College Park, USA.

a)Corresponding author: mdshajedulhoquethakur@ug.me.buet.ac.bd
b)mahmudulridzy@gmail.com, c)nur.jahan123njm@gmail.com, d)bose.buet@gmail.com,
e)adnanmahathir84@gmail.com, f)pial.buet@gmail.com

**Abstract.** Functionally graded materials (FGM) eliminate the stress singularity in the interface between two different materials and therefore have a wide range of applications in high temperature environments such as engines, nuclear reactors, spacecrafts etc. Therefore, it is essential to study the mechanical properties of different FGM materials. This paper aims at establishing a method for modelling FGMs in molecular dynamics (MD) to get a better insight of their mechanical properties. In this study, the mechanical characteristics of *Cu-Ni* FGM nanowires (NW) under uniaxial loading have been investigated using the proposed method through MD simulations. In order to describe the inter-atomic forces and hence predict the properties properly, EAM (Embedded atom model) potential has been used. The nanowire is composed of an alloying constituent in the core and the other constituent graded functionally along the outward radial direction. Simple Linear and Exponential functions have been considered as the functions which defines the grading pattern. The alloying percentage on the surface has been varied from 0% to 50% for both *Cu*-cored and *Ni*-cored nanowires. All the simulations have been carried out at 300 K. The *L/D* ratios are 10.56 and 10.67 for *Cu*-cored and *Ni*-cored NWs, respectively. This study suggests that Ultimate Tensile Stress (*UTS*) and Young's modulus (*E*) increase with increasing surface *Ni* percentage in *Cu*-cored NWs. However, in *Ni*-cored NWs these values decrease with the increase of surface *Cu* percentage. Also, for the same surface percentage of *Ni* in *Cu*-cored NW, the values are higher in linearly graded FGMs than that in exponentially graded FGMs. While in *Ni*-cored NWs, exponentially graded FGM shows higher values of *UTS* and *E* than those in linearly graded FGM. Thus, grading functions and surface percentages can be used as parameters for modulating the mechanical properties of FGM nanowires.

## INTRODUCTION

Understanding the mechanical properties of fabricated materials is a challenging topic in the field of material science. Fabricated materials such as composites comprising of two or more different materials are used for their enhanced strength, efficiency and durability. One of the disadvantages of composite materials is the stress singularity developed at the interfaces between two different materials. In high temperature environment this singularity of stress caused by material mismatch will create higher residual stress which may result in cracking. To eliminate the stress singularity while attaining the enhanced strength and durability of composite materials, the concept of functionally graded material (FGM) has been introduced.[1,2] An FGM is a fabricated material consisting of different material constituents where the constituents' volume fraction continuously varies following a particular function.[3] As there is a gradual change in material volume fraction which in turns results in gradually varying material properties, the stress singularity is not present in case of FGM. Consequently, the residual stresses are significantly relaxed in FGM.[4] As FGM has a wide variety of applications in the field of automotive and aviation industries, space vehicles and reactor vessels, the literature corresponding to Functionally Graded Material's constituent, fracture mechanics and processing methods have increased rapidly. Many studies have been conducted to understand the optimum profile of FGM materials. In these studies, three functions are commonly used to describe the variation of constituents of FGMs. They are Power-law function,[5] Exponential function[6] and Sigmoid function.[7] In the present study, exponential function and linear function which is a derivative of power-law function, have been investigated.

Many research works have been conducted to understand the mechanical properties of FGMs. Karman theory has been applied for large deformation in order to get an analytical solution for FGM plates under transverse mechanical loading by Woo and Meguid.[8] Praveen and Reddy[9] studied the nonlinear thermoelasticity of functionally graded ceramic-metal plates. They have used a plate finite element that accounts for the transverse shear strains along with rotary inertia and moderately large rotations in the Von Karman sense. He et al.[10] used a finite element formulation which is based on classical laminated plate theory to study the vibration control of FGM plates. Feldman and Aboudi[11] investigated elastic bifurcation buckling of FGM plates under compressive loading using micromechanical and structural approaches. These research studies are all based either on theoretical model or finite element methods (FEM) which are based on macroscopic approach. To investigate and understand FGMs from a nanoscale point of view, molecular dynamics (MD) simulations can be a very effective method. Although the mechanical properties of nanowires made of different materials and composites are studied extensively through MD simulations, to the best of author's knowledge, no molecular dynamics study has been conducted to investigate the mechanical properties of FGM nanowires. This is due to the fact that, a proper method of modelling FGMs in MD is not present in the literature.

In the present study, a modeling procedure of functionally graded material (FGM) nanowire for molecular dynamics simulation has been proposed and developed. The surface alloying percentage, grading function and the core of the nanowire are three of the many important factors, which play a significant role in the mechanical properties of an FGM nanowire. Therefore, two grading functions, exponential and linear, have been chosen for this study. The aim of this paper is to investigate the effects of the aforementioned variables on $Cu$-$Ni$ FGM nanowires under the application of tensile loading which is very common for nanostructures. The stress-strain curves have been plotted and from there, the Young's moduli and ultimate tensile stresses are quantified and compared to understand the effects of the alloying constituents and their different surface alloying percentages and different grading functions.

## METHODOLOGY

Two types of FGM NWs are considered in this study: $Cu$-cored NWs with nickel as the graded alloying constituent and $Ni$-cored NWs with copper as the graded alloying constituent. For the $Cu$-cored FGM nanowires copper is the main constituent with nickel functionally alloyed along the outward radial direction. For $Ni$-cored NWs, copper is functionally alloyed in nickel nanowire. Other than their vast applications in various different industries, few reasons for choosing copper-nickel alloys are that below 1085°C, at all alloying percentages $Cu$-$Ni$ forms only a single α-phase; also, in α-phase, $Ni$ substitutes $Cu$ randomly from copper's FCC lattice points.[12] Besides, their nearly identical lattice constants (0.352 nm for $Ni$ and 0.36 nm for $Cu$) reduce complications in structure modelling.

For the alloy modeling, in case of $Cu$ cored NWs, first, the pure $Cu$ nanowire in [1 0 0] orientation is prepared. It is then subdivided into concentric annular cylindrical chunks and in each chunk $Cu$ atoms are randomly replaced with $Ni$ atoms for different weight percentage of $Ni$ in $Cu$, according to the FGM function and the surface alloying percentage. This is similar to the random replacement method previously adopted by Mojumder.[13,14] The opposite is done for the preparation of the $Ni$ cored NWs. In this study, two functions have been considered for the grading: exponential and linear.

The mass fraction of $Ni$ ($Cu$) in $Cu$ ($Ni$) for copper (nickel) cored nanowires, $g(r)$, of the linearly graded FGM is assumed to obey the following function:

$$g(r) = \left(\frac{r}{R}\right) g_{max} \quad (1)$$

while that for the exponentially graded FGM is assumed to follow the function:

$$g(r) = \left(e^{\ln(2)\frac{r}{R}} - 1\right) g_{max} \quad (2)$$

where $r$ is the radial distance from the center of the NW and $R$ is the radius of the NW and $g_{max}$ is the maximum percentage of alloying element which will be at the outer curved surface. The value of $g_{max}$ has been varied from 0% to 50% in this study for both linear and exponential grading as illustrated in Fig. 1(a) and 1(b). Three representative structures of copper cored FGM NWs for each grading are presented in Fig. 1(c) and 1(d).

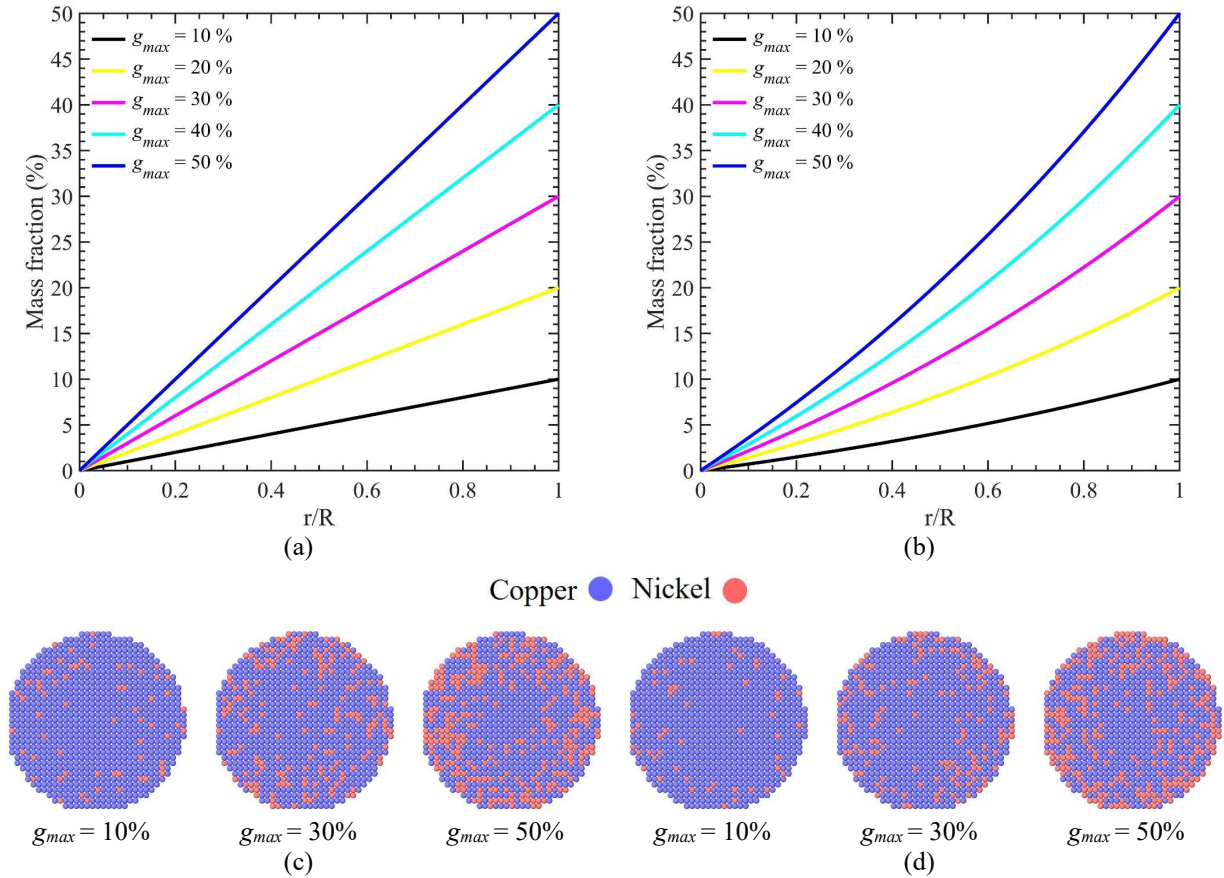

**FIGURE 1.** Variation of mass fraction of alloying constituent along radial direction for (a) linearly graded and (b) exponentially graded FGM NWs for different values of $g_{max}$. The top views of three representative structures for (c) linearly graded and (d) exponentially graded FGM NWs.

The uniaxial tension simulations are performed for *Cu-Ni* FGM nanowires having two different cylindrical structures, their diameters, lengths, aspect ratio and number of atoms are summarized in Table 1. The EAM alloy[15] potential is used to describe the interaction between *Cu* and *Ni* in this study. All the simulations are carried out using LAMMPS[16] software package and OVITO[17] is used for the post-processing purposes.

**TABLE 1.** Geometric specifications of FGM nanowires.

| FGM Structure | Diameter, $D$ (nm) | Length, $L$ (nm) | Number of atoms | $L/D$ |
|---|---|---|---|---|
| *Cu* cored NW | 5.76 | 60.84 | 147030 | 10.56 |
| *Ni* cored NW | 5.81 | 61.95 | 167200 | 10.67 |

At the start of the simulations, the energy of the system is minimized using a conjugate gradient minimization scheme and then the NW is relaxed sufficiently (for 50 ps) using constant NVE integration. In our simulations, the time step has been chosen as 1 fs. The periodic boundary condition is maintained along the loading direction. Then, pressure equilibration has been performed for 100 ps using the isothermal–isobaric (NPT) ensemble at atmospheric pressure and 300 K temperature. The system is then thermally equilibrated using the canonical (NVT) ensemble for 10 ps. A Nose–Hoover thermostat is employed in these steps to maintain the temperature at 300 K. The required timesteps mentioned for NVE, NPT and NVT simulation are chosen by trial and error to equilibrate the various state variables. Finally, uniaxial strain is applied along the *z*-direction at a constant strain rate of $10^9$ s$^{-1}$. This is considered to be a favorable strain rate for MD simulations considering the computational constraints of MD, and it has also been successfully employed in previous literature studies.[18–20] The tensile tests are conducted with the NVT ensemble using a Nose–Hoover thermostat at 300 K. Each deformation simulation is carried out until the failure of the FGM NWs.

The atomic stresses in our simulations are calculated using the virial stress theorem,[21] which can be represented as:

$$\sigma_{virial}(\mathbf{r}) = \frac{1}{\Omega} \sum_i \left( -m_i \dot{\mathbf{u}}_i \otimes \dot{\mathbf{u}}_i + \frac{1}{2} \sum_{j \neq i} \mathbf{r}_{ij} \otimes \mathbf{f}_{ij} \right) \tag{3}$$

In this theorem, the summation is performed over all the atoms occupying a volume $\Omega$, $m_i$ represents the mass of atom *i*, $\mathbf{u}_i$ is its displacement; $\dot{\mathbf{u}}_i = d\mathbf{u}_i/dt$ the velocity; $\mathbf{r}_{ij}$ is the position vector of atom *i* with respect to atom *j*; $\otimes$ is the outer, dyadic or direct tensor product of two vectors; and $\mathbf{f}_{ij}$ is the force on atom *i* due to the pair interaction with atom *j*. Young's modulus is calculated using the elastic region of the stress strain curve. Since, in this work, we limited our study to tensile test and two particular gradings only, vibrational test and exhaustive analysis of effects of different grading functions are left as a future study.[22] The structure files used in the present atomistic study have been generated through nanoHUB.org with LAMMPS Input Structure Generator for Functionally Graded Material (FGM) tool (https://nanohub.org/tools/fgmbuilder).[23]

## RESULTS AND DISCUSSIONS

### Method Validation

To make sure that the simulation process is acceptable, a tensile test has been performed for bulk copper. The experimental value for the elastic modulus of copper is 127 GPa[24] while the MD reported elastic modulus is 148.3 GPa[15]. The elastic modulus obtained from the present method is 147.6 GPa, which is in good agreement. Another tensile test has been performed for bulk nickel and the resulting elastic modulus was found to be 221.38 GPa which is in good agreement with reported MD value of 223.57 GPa[15] and experimental values of about 202 GPa[25].

### Stress–Strain Relationship in *Cu* Cored FGM NWs

The stress–strain curves for different surface alloying percentages in *Cu* cored *Cu-Ni* FGM NWs under uniaxial loading at a strain rate of $10^9$ s$^{-1}$ are shown in Fig. 2(a) and 2(b) for exponentially and linearly graded FGM NWs, respectively. It is observed that for pure *Cu* nanowire the ultimate strength is about 6.0 GPa. Moreover, the sharp decrease of stress at the failure point indicates brittle-type failure of the material. As the surface alloying percentage of *Ni* increases from 0% to 50%, the ultimate tensile strength increases to 7.6 GPa. This hardening behavior at higher surface percentages of *Ni* is also observed for linearly graded nanowires as shown in Fig. 2(b). The hardening behavior due to increase in *Ni* percentage is expected as *Ni* is relatively harder than *Cu* and with the increase in $g_{max}$, total *Ni* percentage increases. It can be observed that all the stress–strain curves follow similar pattern until fracture. The measured values of ultimate stress range from 6 to 7.6 GPa and are distributed around an average value of 6.8 GPa. The decrease in ultimate strength occurs gradually with the decrease in *Ni* percentage. It should be noted that, in case of bulk copper and nickel, Young's modulus and ultimate tensile strength are much higher than our MD results. This is because in case of nanowires, surface energy plays a dominant role in determining the mechanical properties and decreases Young's modulus and ultimate tensile strength significantly.

To get the clear concept of the effects of maximum alloying percentage ($g_{max}$) on the tensile behavior of *Cu*-cored FGM nanowires, the resulting ultimate tensile stress values of the present simulations for both exponentially and linearly graded FGM are plotted in Fig. 2(c). It can be observed that the ultimate tensile stress values are higher for the case of linearly graded FGM NWs than those of the exponentially graded in all cases except in the case of 10% maximum *Ni* percentage. The ultimate tensile stress of *Cu* cored NWs at higher *Ni* percentage can be explained as the ultimate tensile stress of pure *Cu* NW (6 GPa) is less than that of *Ni* (13.5 GPa). Therefore, when the number of *Ni* atoms increases in the structure by replacing *Cu* atoms, with increase in $g_{max}$, the structure strength increases. This result is understandable as increase of *Ni* constituent in *Cu* alloy will direct the material properties more towards pure nickel, which has higher *E* and *UTS*. Our proposed MD model of FGM conforms with it and thus can be applied in other FGM structures.

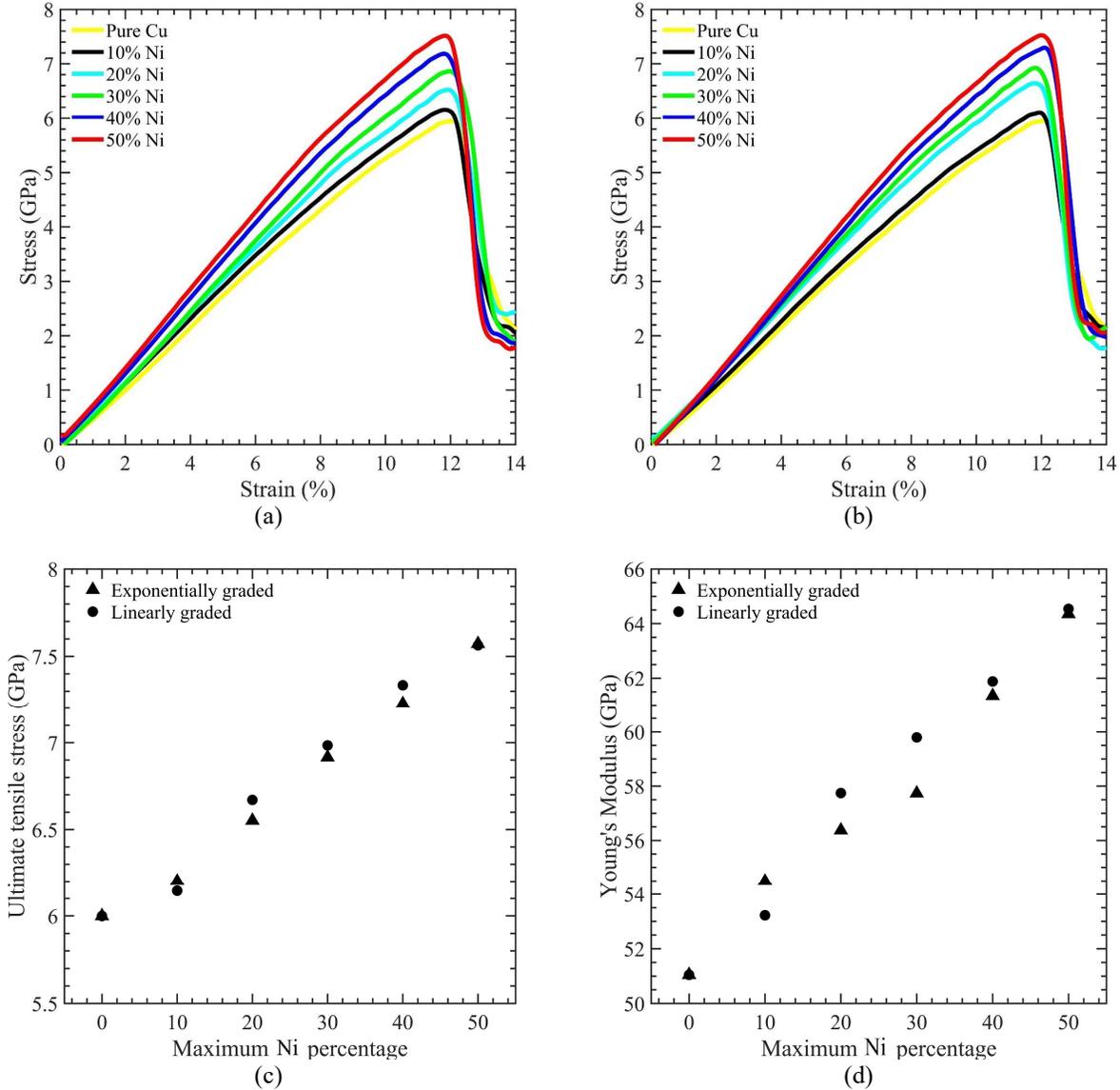

**FIGURE 2.** Stress–strain curves for FGM NWs for various surface alloying percentages of *Ni* with grading: (a) exponential and (b) linear. Variations of the (c) ultimate stress and (d) Young's modulus of exponentially and linearly graded FGM nanowires with maximum *Ni* percentage.

The variation of Young's modulus for exponentially and linearly graded FGM nanowires with maximum *Ni* percentage ($g_{max}$) is presented in Fig 2(d). The results show a nonlinear increasing trend of the young's modulus with the increase of surface alloying percentage of *Ni* in both exponentially and linearly graded FGM nanowires. The Young's moduli were obtained from the stress–strain graphs by fitting the curves up to ultimate stress to a $3^{rd}$ order polynomial equation and the first order coefficient is the young's modulus. Since Young's modulus is the slope of the stress strain curve at zero strain. The Young's moduli of *Cu* cored NWs at higher *Ni* percentage can be explained as the young's modulus of pure *Ni* NW (115 GPa) is greater than that of *Cu* (51 GPa). Therefore, when the number of *Ni* atoms are increasing in the structure by replacing *Cu* atoms, with increase in alloying percentage, the structure becomes stiffer.

# Stress–Strain Relationship in *Ni* Cored FGM NWs

The effects of surface alloying percentage of *Cu* on the stress–strain relationship of *Ni* cored *Cu-Ni* FGM NWs under uniaxial loading are shown in Fig. 3 for a strain rate of $10^9$ s$^{-1}$. In Fig. 3(a) & (b), the effects of alloying percentage are shown for both exponentially and linearly graded *Cu-Ni* FGM NW. It can be inferred from Fig. 3(a) that as the surface alloying percentage of *Cu* rises from 0% to 50%, the ultimate stress decreases from about 13.5 GPa to 10.9 GPa. The stress–strain curves also show a brittle failure pattern for these FGM NWs. This softening behavior at higher surface percentage of *Cu* was also observed for linearly graded nanowires in Fig. 3(b). This softening behavior can be explained as the ultimate tensile stress of pure *Cu* NW (6 GPa) is less than that of *Ni* (13.5 GPa). Therefore, when the number of *Cu* atoms are increasing in the structure by replacing *Ni* atoms, with increase in alloying percentage, the structure loses its strength. It can be observed that all the stress–strain curves follow almost the same path until fracture. The measured values of ultimate stress range from 13.5 to 10.4 GPa and are distributed around an average value of 11.95 GPa.

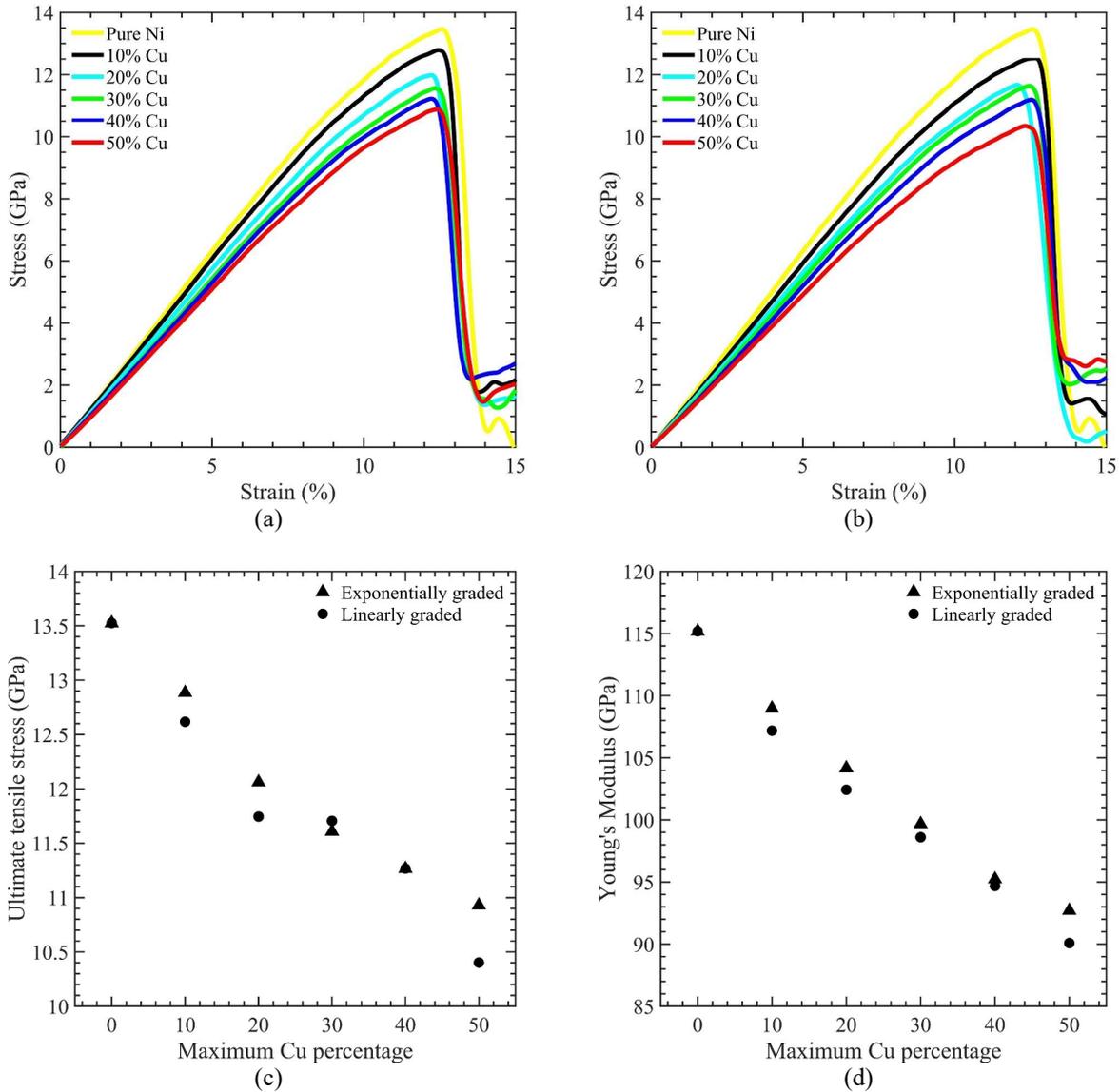

**FIGURE 3.** Stress–strain curves for FGM NWs for various surface alloying percentages of *Cu* with grading: (a) exponential and (b) linear. Variations of the (c) ultimate stress and (d) Young's modulus of exponentially and linearly graded FGM nanowires with maximum *Cu* percentage.

To observe the influence of different functional grading on the fracture behavior of *Cu*-cored *Cu-Ni* FGM NWs, the ultimate tensile stress values for both exponentially and linearly graded FGM NWs are plotted against the surface alloying percentage of *Cu* in Fig. 3(c). The plotted results in Fig. 3(c) show that the ultimate stress vs *Cu* percentage curves follow a similar but decreasing path, unlike the increasing paths of their *Cu*-cored counterparts. The ultimate stress values in exponentially graded FGM are slightly greater than linearly graded FGM at same alloying percentage and the deviation increasing up to 0.5 GPa.

The variations of the Young's modulus with alloying percentage for both exponentially and linearly graded *Ni*-cored FGM are presented in Fig. 3(d). This plot shows exactly the opposite trends that were observed in the case of *Cu*-cored FGM NWs. Although the nature of the graphs is the same, the results may provide more insight when they are seen quantitatively. Here the deviation of the young's modulus is maximum 2 GPa. *Ni*-cored FGM nanowires show about a 79.4% increment of the average ultimate stress for *Ni*-cored FGM compared to their *Cu*-cored counterparts. Moreover, the average value of the Young's modulus is also about 80% higher in the *Ni*-cored FGM NWs. Therefore, it can be concluded that *Ni*-cored *Cu-Ni* FGM NWs are stronger and stiffer in nature than *Cu* cored *Cu-Ni* FGM NWs because of the presence of *Ni* atoms in high amount.

One interesting result observed in the present work is that, in *Cu*-cored NWs, *E* and *UTS* of linearly graded NWs are greater than exponentially graded NWs. On the other hand, in case of *Ni*-cored NWs, *E* and *UTS* of exponentially graded NWs are higher than linearly graded NWs. This can be explained by Fig. 1(a) and Fig. 1(b). The area under the curves of both Fig. 1(a) and Fig. 1(b), represent the total *Ni* constituent in case of *Cu*-cored NW and total *Cu* constituent in case of *Ni*-cored NW. And for all the values of $g_{max}$, the area is always greater in linear grading compared to exponential grading. Hence in *Cu*-cored NWs, for a particular value of $g_{max}$, *Ni* constituent is always greater in linear grading and thus its *E* and *UTS* are greater than exponential grading. Same argument is applicable in case of *Ni*-cored NWs.

## CONCLUSIONS

An atomistic modelling method for functionally graded material (FGM) has been proposed and developed for MD simulation. Using this model, two types of radially functionally graded nanowires comprising of copper and nickel having a wide range of surface nickel percentages have been generated. Then tensile tests have been conducted through molecular dynamics simulations. For proper comparison pure copper and pure nickel nanowires have been considered along with five copper cored functionally graded nanowires (Surface nickel percentages of 10% to 50%) and five nickel cored functionally graded nanowires (surface copper percentages of 10% to 50%). To describe the interatomic interactions between the atoms present in simulations, EAM alloy potential has been used. The results have been presented through stress-strain curves, elastic modulus and ultimate tensile strength. The conclusive results are following:

1. Young's modulus and ultimate tensile strength of linearly graded *Cu*-cored NWs are greater than exponentially graded NWs.
2. In case of *Ni*-cored NWs, Young's modulus and ultimate tensile strength of exponentially graded NWs are greater than linearly graded NWs
3. With the increase of surface nickel percentages, elastic modulus increases gradually to reach 64.5 GPa at 50% surface nickel percentage, in case of copper cored nanowires.
4. As the surface copper percentage increases, elastic modulus decreases gradually to reach 90 GPa at 50% surface copper percentages, in case of nickel cored nanowires.
5. The ultimate stress increases with increasing nickel percentages and decrease with increasing surface copper percentages.
6. The present method of modelling FGMs in MD can be applied in other geometric FGM structures.

## ACKNOWLEDGMENTS

The authors of this paper would like to acknowledge Multiscale Mechanical Modeling and Research Networks (MMMRN) for their technical assistance and especially Satyajit Mojumder, Northwestern University, for providing the computational facilities to conduct the research. We also acknowledge the contribution of Professor Md Mahbubul Islam, Wayne State University and Abdullah Al Amin, Bridgestone, for their insightful comments on the paper.